\documentclass[11pt]{article}
\usepackage{amsmath, amssymb,amsthm}
\usepackage{graphics,color,booktabs}
\usepackage{easybmat}
\usepackage{multirow,threeparttable}
\renewcommand{\baselinestretch}{1.5}

\numberwithin{equation}{section}

\addtocounter{page}{0}
\newtheorem{thm}{Theorem}[section]

\newtheorem{lem}{Lemma}[section]
\allowdisplaybreaks

\begin{document}

\title{\bf Penalized Maximum Likelihood Estimator for Skew Normal Mixtures}
\author{Libin Jin, Wangli Xu, Liping Zhu and Lixing Zhu\footnote{ Libin, Jin is a PHD student of School of Statistics at Renmin University of China, Beijing, China, Wangli Xu is a Professor of School of Statistics at Renmin University of China, Beijing, China Liping Zhu is a Professor of Institute of Statistics and Big Data at Renmin University of China, Beijing, China, Lixing Zhu is a Chair Professor of Department of Mathematics at
Hong Kong Baptist University, Hong Kong, China, and a Professor of School
of Statistics at Beijing Normal University. The corresponding email is lzhu$@$hkbu.edu.hk He was supported by a grant from
the University Grants Council of Hong Kong, Hong Kong, China, and a grant from
the National Science Foundation of China.}
%1. School of Statistics, Renmin University of China, Beijing, China; \\
%2. Institute of Statistics and Big Data, Renmin University of China, Beijing, China;
% \\
%3. School of Statistics, Beijing Normal University, Beijing, China.\\
%4. Department of Mathematics, Hong Kong Baptist University, Hong Kong, China.
}
\maketitle

\begin{abstract}
Skew normal mixture models provide a more flexible  framework
than the popular normal mixtures for modelling heterogeneous data
with asymmetric behaviors. Due to the unboundedness of likelihood
function and the divergency of shape parameters, the maximum likelihood
estimators of the parameters of interest are often not well defined, leading to dissatisfactory inferential process.
We put forward a proposal to deal with these issues simultaneously
in the context of penalizing the likelihood function.
The resulting penalized maximum likelihood estimator is proved to be  strongly consistent when the putative order of mixture is equal to
or larger than the true one. We also provide penalized EM-type algorithms to compute  penalized estimators. Finite sample performances are examined by
simulations and  real data applications and the comparison to the  existing methods.

\end{abstract}

%\footnote{ Libin, Jin is a PHD student of School of Statistics at Renmin University of China, Beijing, China, Wangli Xu is a Professor of School of Statistics at Renmin University of China, Beijing, China Liping Zhu is a Professor of Institute of Statistics and Big Data at Renmin University of China, Beijing, China, Lixing Zhu is a Chair Professor of Department of Mathematics at
%Hong Kong Baptist University, Hong Kong, China, and a Professor of School
%of Statistics at Beijing Normal University. The corresponding email is lzhu\@hkbu.edu.hk He was supported by a grant from
%the University Grants Council of Hong Kong, Hong Kong, China, and a grant from
%the National Science Foundation of China.}

\section{Introduction}

Finite mixtures of skew normal (SN) distributions have
received considerable attention in recent years.
In tackling data with multimodal and asymmetric behaviours,
skew normal mixture (SNMIX) models are considered as a more flexible
and robust tool than the most popular Gaussian mixture(GMIX) models.
With component densities themselves capturing skewness and excess kurtosis,
this framework  remedies unrealistic symmetric assumptions
and avoids the overfitting problem existing in GMIX
(Lin {\em et al.} 2007b; Fruhwirth-Schnatter and Pyne 2010).

The several attempts to analyse skew normal mixtures are attributed to
Lin {\em et al.} (2007b) and Basso {\em et al.} (2010).
A multivariate extension of this model has been developed
by Lin (2009) and Cabral {\em et al.} (2012).
Fruhwirth-Schnatter and Pyne (2010) explored a Bayesian approach and
proposed an efficient MCMC scheme in multivariate SNMIX.
Other researchers extended the SN distribution
to more general statistical models,
such as linear mixed models (Lachos {\em et al.} 2010),
parsimonious clustering models (Vrbik and McNicholas 2014) and
mixtures of regression models (Zeller {\em et al.} 2016).

Consider the SN distribution introduced by Azzalini (1985),
whose density function is given by
\begin{equation}\label{snp}
f_{SN}(x;\theta) = \frac{2}{\sigma} \phi\bigg(\frac{x-\mu}{\sigma}\bigg)
\Phi\bigg(\lambda\frac{x-\mu}{\sigma}\bigg), \ x\in\mathbb{R}
\end{equation}
where $\theta=(\mu,\sigma^2,\lambda)\in\Theta\subseteq\mathbb{R}\times\mathbb{R}^+\times\mathbb{R}$,
and $\phi(\cdot)$ and $\Phi(\cdot)$ are the normal density and distribution function.
The density (\ref{snp}) depends on $\mu,\sigma^2,\lambda$,
which regulate location, scale and shape (skewness) respectively.

Given the kernel density (\ref{snp}) and a finite order $p$,
as shown in Lin {\em et al.} (2007b),
the density function of SNMIX is
\begin{equation}\label{snmp}
f(x;\Psi)=\sum_{k=1}^p \pi_k f_{SN}(x;\theta_k)=\int f_{SN}(x;\theta)d\Psi(\theta)
\end{equation}
where $\pi_k,\theta_k=(\mu_k,\sigma^2_k,\lambda_k)$ are the mixing proportion
and component parameters respectively.
We use the notation $\Psi$ for all parameters in SNMIX,
and for its cumulative distribution function
$\Psi(\theta)=\sum_{k=1}^p\pi_kI(\theta_k\leq\theta)$,
where $I(\cdot)$ is the indicator function.

The parameter space of $\Psi$ can then be written as
\begin{align*}
\Gamma= \bigg\{ &\Psi=(\pi_1,\cdots,\pi_p,\mu_1,\cdots,\mu_p,
\sigma_1,\cdots,\sigma_p,\lambda_1,\cdots,\lambda_p): \\
 & 0\leq\pi_k\leq1,\sum_{k=1}^p\pi_k=1,-\infty<\mu_k,\lambda_k<+\infty,
\sigma_k\geq0,k=1,\cdots,p \bigg\}.
\end{align*}

In finite mixture models, several approaches are available in the literature,  which characterize the mixing distribution $\Psi$, see Lindsay (1995)
and McLachlan \& Peel (2000). Among which,
the maximum likelihood estimator (MLE) is commonly used for
its asymptotic efficiency under regular parametric models.
In univariate and multivariate SNMIX,
Lin {\em et al.} (2007b) and Lin (2009)  investigated the theory and
applications of the MLE as well as corresponding EM algorithms.

However, the ordinary MLE may be not well defined even in  the classical normal mixtures
(Kiefer and Wolfowitz 1956; Day 1969).
Suppose we have a random sample $\{X_1,\cdots,X_n\}$ of size $n$
from the above SNMIX model. Then the log-likelihood function is
\begin{equation}\label{snll}
\ell_n(\Psi)=\sum_{i=1}^n\log f(X_i;\Psi)=\sum_{i=1}^n\log\bigg\{
\sum_{k=1}^p\frac{2\pi_k}{\sigma_k} \phi\bigg(\frac{X_i-\mu_k}{\sigma_k}\bigg)
\Phi\bigg(\lambda_k\frac{X_i-\mu_k}{\sigma_k}\bigg) \bigg\}
\end{equation}
It is clear that $\ell_n(\Psi)$ is unbounded over
parameter space $\Gamma$ for any given $n$,
due to it goes to infinity as $\mu_k \rightarrow X_i$ and
$\sigma_k \rightarrow 0$ with the other parameters fixed (Ciuperca {\em et al.} 2003).
Hence, a global MLE of $\Psi$ is known to be inconsistent.
Meanwhile, the likelihood ratio test statistic
is shown to lose the elegant asymptotic properties.

To avoid likelihood degeneracy, two likelihood-based approaches were proposed
to regain the consistency and efficiency. One is the constraint MLE.
Redner (1981) proved that, in every compact parameter subspace
containing the true parameter $\Psi_0$,
the MLE $\hat\Psi\rightarrow\Psi_0$ in probability as $n\rightarrow\infty$.
Hathaway (1985) suggested  using a constrained MLE under the condition
$\underset{i,j}{\min}\ \sigma_i/\sigma_j\geq c>0$,
where $c$ is a fixed constant. However, as stated in Chen {\em et al.} (2008),
the reduction of parameter space may lead to the true parameter $\Psi_0$
not belonging to the altered space.
Other researchers focused on the penalized method.
It is a promising approach  to counter the likelihood unboundedness problem
without altering the parameter space.
With different penalties on component variances,
Ciuperca {\em et al.} (2003) and Chen {\em et al.} (2008) respectively
proved the strong consistency of the penalized maximum
likelihood estimators (PMLE).

In addition to the unbounded likelihood, another
undesirable property in SNMIX is that the MLE of $\lambda_k$ diverges.
For the SN distribution, Azzalini and Capitanio (1999) found that
the MLE of $\lambda$ can occur on the boundary (i.e. $\hat\lambda=\pm\infty$),
even for data whose distribution can be fairly well
approximated by the SN model with finite $\lambda$.
To obtain reliable estimators of $\lambda$, the larger sample sizes are often required
(DiCiccio and Monti 2004).
Azzalini and Arellano-Valle (2013) proved that
$\underset{n\rightarrow\infty}{\lim}P(|\hat\lambda|\rightarrow\infty)=0$,
but the divergence of  MLE occurs with a non-negligible probability for finite sample size.
In SNMIX, for $\Phi(\cdot)$ being a monotonically increasing function,
$\ell_n(\Psi)$ in (\ref{snll}) over $\Gamma$ is maximized at
$$
\hat\lambda_k=\bigg\{
\begin{array}{cc}
\infty,  &  \sum_{i=1}^nI(X_i>\mu_k)=n \\
-\infty, &  \sum_{i=1}^nI(X_i<\mu_k)=n
\end{array}
$$
Although $\hat\lambda_k=\pm\infty$ will not lead to degenerate likelihood,
the standard asymptotic distribution theory of  MLE does not hold
on the boundary of $\Gamma$.
Furthermore, an divergent estimator requires an enormous amount of computational workload
and has unpleasant effects on inferential process (Azzalini and Arellano-Valle 2013).

Unfortunately, under this peculiar situation, the  constraint MLE has no way to alleviate  the divergency of shape parameter estimation.
As an example, place an additional constraint $\max_k\{ |\lambda_k| \}\leq C$
on $\Gamma$, where $C$ is a sufficiently large positive constant.
$\ell_n(\Psi)$ would be maximized only if shape parameters converge to
the boundary of the constrained $\Gamma$ as
\begin{equation}\label{sncml}
\max_k|\hat\lambda_k|=C, \ \textrm{if} \ \max_{k}\bigg|\sum_{i=1}^n \textrm{sgn}(X_i-\mu_k)\bigg|=n
\end{equation}
where $\textrm{sgn}(\cdot)$ is the sign function.
Therefore, the constrained MLE turns out to be invalid.

In this paper, to overcome both likelihood degeneracy and divergent shape parameters,
we recommend estimating $\Psi$ by maximizing the likelihood function
with a penalty function. The penalized log-likelihood is defined as
\begin{align}\label{snpl}
\begin{split}
&p\ell_n(\Psi)=\ell_n(\Psi)+p_n(\Psi), \\
&p_n(\Psi)=\sum_{k=1}^pp_n(\sigma_k)+\sum_{k=1}^pp_n(\lambda_k).
\end{split}
\end{align}
Then PMLE   of $\Psi$ would be obtained by
$\tilde\Psi = \arg\max_{\Psi}p\ell_n(\Psi)$. With reasonable penalties,
the corresponding penalized likelihood $p\ell_n(\Psi)$ is bounded over $\Gamma$,
granting the existence of PMLE.
To regain the consistency of $\tilde\Psi$, penalty functions
$p_n(\sigma)$ and $p_n(\lambda)$ must be chosen carefully.
We select $p_n(\sigma)$ such that it goes to negative infinity
when $\sigma$ goes to either 0 or infinity,
and choose $p_n(\lambda)$ such that $p_n(\lambda)$ tends to negative infinity
as $|\lambda|$ tends to infinity.

We focus on investigating the penalized likelihood-based estimator in skew normal mixtures.
The remainder of the article unfolds as follows.
Section~2 outlines some preliminaries including
technical lemmas and choice of penalties.
In Section~3, we provide a rigorous proof of the strong consistency
of the proposed PMLE in both  $p=p_0$ and $p>p_0$ cases.
The penalized EM algorithms are presented in Section~4.
The simulation results as well as two application examples are
respectively in Section~5 and 6.
Technical proofs are relegated to the Appendix.

\section{Preliminaries}

\subsection{Technical lemmas}

In normal mixture models, Chen {\em et al.} (2008) provided
a novel technique to establish the strong consistency.
Based on the Bernstein Inequality, they proved an insightful conclusion,
the number of observations falling in a small neighbourhood of
the location parameters has a uniform upper bound.
However, as noted in Chen {\em et al.} (2008), the normality assumption does not
play a crucial role. The conclusion has recently been furthered
to the distribution-free case by Chen (2016). Without proofs,
we conclude the main results as the following Lemma \ref{nn}.

\begin{lem}\label{nn}
Let $X_1,\cdots,X_n$ be i.i.d. observations from an absolute continuous distribution
$F$ with density function $f(x)$. Suppose $f(x)$ is continuous and $M=\sup_x f(x)<\infty$.
Let $F_n(x)=n^{-1}\sum_{i=1}^nI(X_i\leq x)$ be the empirical distribution function.
Thus, as $n\rightarrow\infty$,
\begin{equation*}
\underset{x\in\mathbb{R}}{\sup} \{ F_n(x+\epsilon)- F_n(x)\}\leq 2M\epsilon +10n^{-1}\log n,
\end{equation*}
holds uniformly for all $\epsilon>0$ almost surely.
\end{lem}

It is worth observing that, Lemma \ref{nn} excludes the zero-probability event
for each $\epsilon$ on which the upper bound is violated.
Furthermore, it is clear that the density and distribution function
of skew normal mixtures satisfy the milder distribution assumptions in Lemma \ref{nn}.
Thus, let $\epsilon=|\sigma\log\sigma|$, where $\sigma>0$ and $\sigma$ is small.
With a slight alteration, we state the conclusion for skew normal mixtures as follows:

\begin{lem}\label{nn2}
Suppose $X_i,i=1,\cdots,n$ are i.i.d. random samples from the finite mixture of
skew normal distributions with density function $f(x;\Psi_0)$ as defined in (\ref{snmp}),
except for a zero-probability event not depending on $\sigma$, we have
$$
\underset{\mu\in\mathbb{R}}{\sup}\sum_{i=1}^n I\left(|X_i-\mu|\leq|\sigma\log\sigma|\right) \leq
4Mn|\sigma\log\sigma|+10\log n,\ a.s. \ \ \textrm{as} \ \ n\rightarrow\infty.
$$
in which $M=\sup_x f(x;\Psi_0)$.
\end{lem}

\textit{Remark}. For $n\rightarrow\infty$ much faster than $\log n$,
the first item dominates the upper bound.

\subsection{Choice of penalties}

Lemma \ref{nn} and \ref{nn2} provide a technical basis for the sizes of the penalties.
To ensure the consistency of the proposed PMLE,
we assume the following conditions on $p_n(\sigma)$ and $p_n(\lambda)$:
\begin{itemize}
\item [$\mathbf{C1}$.] $\forall\sigma>0$, $p_n(\sigma)=o(n)$ and
$\sup_{\sigma>0}\max\{ 0,p_n(\sigma) \}=o(n)$.
\item [$\mathbf{C2}$.] $p_n(\sigma)\leq (\log n)^2\log\sigma$,
when $\sigma<n^{-1}\log n$ and $n$ is large.
\item [$\mathbf{C3}$.] $p_n(\lambda)$ is a continuous function that takes maximum at $\lambda=0$ and goes to negative infinity
as $|\lambda|\rightarrow\infty$. Besides, $p_n(0)=0$.
\item [$\mathbf{C4}$.] $p_n(\sigma)$ and $p_n(\lambda)$ are differentiable
with respect to $\sigma$ and $\lambda$ respectively, and as $n\rightarrow\infty$,
$p'_n(\sigma)=o(n^{1/2})$ and $p'_n(\lambda)=o(n^{1/2})$.
\end{itemize}

However, the existence of the above required penalty functions is obvious and of non-uniqueness.
Users therefore have the freedom to choose  penalties,
indicating the added mathematical conditions are not restrictive.
Condition C1 makes a restriction on the upper and lower bounds of $p_n(\sigma)$,
while C2 makes $p_n(\sigma)$ sufficiently severe to prevent $\sigma^2\rightarrow 0$.
Condition C3 limits the effect of $p_n(\lambda)$.
Condition C4 guarantees the existence of a limiting distribution of the penalized MLE.
Here, with sample variance denoted by $s^2_n$, we recommend to use
the following two penalty functions
\begin{align}\label{cop}
\begin{split}
&p_n(\sigma) = -a_n\left(s^2_n/\sigma^{2}+\log(\sigma^2/s^2_n)-1\right), \\
&p_n(\lambda) = -b_n\left(\lambda^2-\log(1+\lambda^2)\right).
\end{split}
\end{align}
where $a_n$ and $b_n$ are positive tuning parameters of $p_n(\sigma), p_n(\lambda)$ respectively.

Note that Conditions C1-C4
are easy to verify for the recommended penalties.
The form of $p_n(\sigma)$ also stands for a prior inverse Gamma distribution
placed on $\sigma^2$ from the Bayesian point of view and has the advantage
of retaining scale invariance (Chen {\em et al.} 2008).
It is also well used in constructing EM-test statistic,
see Chen and Li (2009) and Chen {\em et al.} (2012).

The penalty function $p_n(\lambda)$ in (\ref{cop}),
compared with $p(\lambda)=-c_1\log(1+c_2\lambda^2)$
used in Azzalini and Arellano-Valle (2013),
in which $c_1$ and $c_2$ are two fixed constants, has several markedly advantages.
Firstly, as a convex function, $p_n(\lambda)$ is fairly flat near zero and very steep
when $\lambda$ is away from 0. Hence, it has little effects on likelihood function
when $\lambda$ is regular, while sensitive to the divergent skewness parameter.
Furthermore, it is also remarkable that $p_n(\lambda)$ will not
increase computation complexity in the EM-type algorithms.

The sensible choice of $a_n$ and $b_n$ should depend on $n$.
 Under large sample case as $n\to \infty$, Chen {\em et al.} (2012) pointed out that the asymptotic property
of the EM-test statistic will not be changed whenever $a_n=o(n^{1/4})$.
The consistency of PMLE can also be granted whenever $b_n=o(n)$.
In practice, we recommend
\begin{equation}\label{tp}
a_n=c_a/n,b_n=c_b/\log n
\end{equation}
in which the constants $c_a$ and $c_b$ control the scale of penalties.
In this paper, we take $c_a=1$ and $c_b=0.05$. Chen et al (2012) is a reference that also took $c_a=1$.

\section{Strong Consistency of The Penalized MLE}

\subsection{Consistency of The Penalized MLE when $p=p_0$}

Let $K_0=E_{\Psi_0}(\log f(X;\Psi_0))$ denote conditional expectation
under the true mixing distribution and recall $M=\sup_x f(x;\Psi_0)$ in Lemma \ref{nn2}.
Suppose that $\epsilon_0$ and $\eta_0$ are sufficiently small
and large positive constant respectively.
Given $p, M$ and $K_0$, there exists $\epsilon_0\rightarrow0$
satisfying following two inequalities:
$$
4pM\epsilon_{0}\log^2\epsilon_{0}\leq1 \ \textrm{and} \
\log\epsilon_{0}+\frac{\log^2\epsilon_{0}}{2}\geq p(2-K_0).
$$
Besides, we also select a $\eta_0$ such that
$\eta_0>\max_k\{ |\lambda_{0k}| \},k=1,\cdots,p$,
where $\lambda_{0k}$ is the element of $\Psi_0$.
The choice of $\epsilon_0$ and $\eta_0$ clearly depend on $\Psi_0$
but not on the sample size $n$.

For the obvious existence of $\epsilon_0$ and $\eta_0$,
it is convenient to define regions:
\begin{align*}
&\Gamma_{\sigma}=\{ \Psi\in\Gamma: \min\{ \sigma_k\}\leq\epsilon_0, k=1,\cdots,p \}, \\
&\Gamma_{\lambda}=\{ \Psi\in\Gamma: \max\{ |\lambda_k|\}\geq\eta_0, k=1,\cdots,p \}, \\
&\Gamma^*=\Gamma-\Gamma_{\sigma}\cup\Gamma_{\lambda}.
\end{align*}

We will see that the penalization will be on these regions of the parameters. When a vector is in the region $\Gamma_{\sigma}$, then the parameter of the mixing distribution has at least
one component deviation close to zero.
The penalty $p_n(\sigma)$ will counter it such that PMLE with $\sigma \in \Gamma_{\sigma}$ is with a diminishing probability. Similarly, PMLE will exclude the values in
 the region $\Gamma_{\lambda}$ in which there is at east
one $|\lambda_k|$ diverges to infinity.
%which would be prevented by the penalty $p_n(\lambda)$
%so that almost surely $\tilde{\Psi}\notin\Gamma_{\lambda}$.

We first give the consistency in the following theorem. %prove that, regardless of the divergence of skew parameters,
%the probability $P(\tilde{\Psi}\in\Gamma_{\sigma})=0$ as $n\rightarrow\infty$.
%Then it is convenient to discuss the case of $\Psi\in\Gamma^c_{\sigma}\cap\Gamma_{\lambda}$.
To state the results clearly, rearrange the component deviations in ascending order as $\sigma_{(1)}\leq\cdots\leq\sigma_{(p)}$,
with the corresponding mixing proportion and parameters being respectively denoted as
$\pi_{(k)}$ and $\theta_{(k)}=(\mu_{(k)},\sigma^2_{(k)},\lambda_{(k)})$ when $k\in\{1,\cdots,p\}$.
Hence, for $\tau\in\{1,\cdots,p\}$, the parameter space $\Gamma_{\sigma}$ can be partitioned by
$$
\Gamma_{\sigma}^{\tau}=\{ \Psi\in\Gamma_{\sigma}:
\sigma_{(1)}\leq\cdots\leq\sigma_{(\tau)}\leq\tau_{0}
<\epsilon_{0}\leq\sigma_{(\tau+1)}\leq\cdots\leq\sigma_{(p)} \}.
$$
In particular, when $\tau=p$,
$$
\Gamma_{\sigma}^{p}=\{ \Psi\in\Gamma_{\sigma}:
\sigma_{(1)}\leq\cdots\leq\sigma_{(p)}\leq\tau_{0}
<\epsilon_{0} \}.
$$

\begin{thm}\label{thm1}
Assume that the density function  is $f(x;\Psi_0)$. Let the penalized likelihood  $p\ell_n(\Psi)$ be defined as in (\ref{snpl})
with the penalty function $p_n(\Psi)$ satisfying C1-C3.
Then for any $\Psi\in\Gamma_{\sigma}^p$, as $n\rightarrow \infty$ and almost surely
$$
\sup_{\Gamma_{\sigma}^p} p\ell_n(\Psi)-p\ell_n(\Psi_0)\rightarrow -\infty.
$$
\end{thm}
The proof of Theorem  \ref{thm1} is in Appendix.
For the spaces $\Gamma_{\sigma}^\tau$ with $1\leq\tau\leq p-1$,
we can obtain the similar results as in Theorem \ref{thm1}. The result is stated below.

\begin{thm}\label{thm2}
Under the same assumptions as in Theorem \ref{thm1}
except that $\Psi\in\Gamma_{\sigma}^\tau$ for $\tau$ with $1\leq\tau\leq p-1$,
$\sup_{\Psi\in\Gamma_{\sigma}^\tau} p\ell_n(\Psi)-p\ell_n(\Psi_0)\rightarrow -\infty$
almost surely as $n\rightarrow \infty$.
\end{thm}

Note that $\Gamma_{\sigma}=\cup_{\tau=1}^{p}\Gamma_{\sigma}^{\tau}$.
 From Theorems \ref{thm1} and \ref{thm2}, we conclude that PMLE of $\Psi$ is not in $\Gamma_{\sigma}$
except for a zero probability event.
Below, we present a result showing the boundedness of skewness parameters $\lambda$ in PMLE. Consider the region $\Psi\in\Gamma^c_{\sigma}\cap\Gamma_{\lambda}$.

\begin{thm}\label{thm3}
Under the same conditions as in Theorem \ref{thm1}, as $n\rightarrow \infty$,
we can also show that almost surely
$\sup_{\Psi\in\Gamma^c_{\sigma}\cap\Gamma_{\lambda}} p\ell_n(\Psi)-p\ell_n(\Psi_0)\rightarrow -\infty$.
\end{thm}

From the above three Theorems, we have exclude the possibility that
the penalized MLE $\tilde{\Psi}$ falls in $\Gamma_{\sigma}\cup\Gamma_{\lambda}=\Gamma_{\sigma}\cup\{\Gamma^c_{\sigma}\cap\Gamma_{\lambda}\}$.
Hence, it suffices to show that $\tilde{\Psi}\in\Gamma^*$ with probability $1$.
The strong consistence of $\tilde{\Psi}$ is stated below.

\begin{thm}\label{thm4}
Assume the same conditions as in Theorem \ref{thm1},
$\Psi$ is a mixing distribution with $p_0$ components satisfying
$$
 p\ell_n(\Psi)-p\ell_n(\Psi_0)\geq c > -\infty.
$$
Then as $n\rightarrow \infty$, $\Psi\rightarrow\Psi_0$ almost surely.
\end{thm}

Rewrite $\Gamma^*=\{ \Psi\in\Gamma: \min_k\{ \sigma_k\}\geq\epsilon_0,\max_k\{ |\lambda_k|\}\leq\eta_0 \}$,
$\Psi\in\Gamma^*$ is equivalent to impose a positive lower bound to component deviations
and a positive upper bound to the absolute value of skewness parameters.
Since $\Gamma^*$ is regular, the consistency is then covered by the technique
in Kiefer and Wolfowitz (1956) even with a penalty of size $o(n)$.

Since $ p\ell_n(\tilde{\Psi})-p\ell_n(\Psi_0)\geq 0$,
the PMLE $\tilde{\Psi}$ is thus strongly consistent.
Besides, for $p=p_0$, all elements in $\tilde{\Psi}$
converge to those of $\Psi_0$ almost surely.

Further, let
$S_n(\Psi)=\frac{\partial\ell_n(\Psi)}{\partial\Psi}$ and
$S'_n(\Psi)=\frac{\partial^2\ell_n(\Psi)}{\partial\Psi\partial\Psi^T})$
be respectively the score vector and second derivative matrix of $\ell_n(\Psi)$.
Since the SNMIX model is regular at $\Psi_0$,
we have the positive definite fisher information matrix
$I(\Psi_0)=-E\{S'_n(\Psi_0)\}=E\{S^T_n(\Psi_0)S_n(\Psi_0)\}$.
Based on the classical asymptotic technique
and condition C4 such that
$p'_n(\sigma)=o(n^{1/2}),\, p'_n(\lambda)=o(n^{1/2})$,
we have
$$
\tilde{\Psi}-\Psi_0=-\{S'_n(\Psi_0)\}^{-1}S_n(\Psi_0)+o_p(n^{1/2}).
$$
Thus, the penalized estimator is of the asymptotic normality and efficiency.
\begin{thm}\label{thm5}
Under the same conditions as in Theorem \ref{thm1} and Condition C4,
as $n\rightarrow\infty$
$$
\sqrt{n}(\tilde{\Psi}-\Psi_0)\rightarrow N(\mathbf{0},I^{-1}(\Psi_0))
$$
in distribution.
\end{thm}

\subsection{Consistency of The Penalized MLE when $p>p_0$}

In practice, it is often that people only  know an upper bound of
the mixture order rather than the exact $p_0$, that is, $p_0<p<\infty$.
In this case, by treating both $\tilde{\Psi}$ and $\Psi_0$ as mixing
distributions on the same space, Chen {\em et al.} (2008) and Chen and Tan (2009)
 proved the consistency of their PMLEs in univariate
and multivariate normal mixtures. To measure the difference
between the mixing distributions $\Psi$ and $\Psi_0$,
we first define a distance as
\begin{equation}\label{dst}
D(\Psi,\Psi_0)=\int_{\Theta} |\Psi(\theta)-\Psi_0(\theta)|\exp(-|\theta|)d\theta
\end{equation}
where $\theta=(\mu,\sigma^2,\lambda)\in\Theta\subseteq\mathbb{R}\times\mathbb{R}^+\times\mathbb{R}$,
$|\theta|=|\mu|+\sigma^2+|\lambda|$ and $d\theta=d\mu d\sigma^2 d\lambda$.
The distance has two desirable properties. First, it is bounded  with the inequalities
$0\leq D(\Psi,\Psi_0)\leq\int_{\Theta} \exp(-|\theta|)d\theta<\infty$.
Second, $D(\tilde{\Psi},\Psi_0)\rightarrow0$ implies that $\tilde{\Psi}\rightarrow\Psi_0$
in distribution, providing the technical basis for consistency.
Hence, we have the following theorem.

\begin{thm}\label{thm6}
Assume the same conditions as in Theorem \ref{thm1}, except that $p_0< p<\infty$,
for any mixing distribution $\Psi$ with $p$ components satisfying
$$
 p\ell_n(\Psi)-p\ell_n(\Psi_0)\geq c > -\infty.
$$
Then as $n\rightarrow \infty$,  $\Psi\rightarrow\Psi_0$ almost surely.
\end{thm}

\section{Penalized EM Algorithms}

Concerning  computation,
Lin {\em et al.} (2007b) exploited two extensions of the EM algorithm:
the ECM algorithm (Meng and Rubin 1993) and the ECME algorithm (Liu and Rubin 1994).
In view of the asymptotic  properties (Hero and Fessler 1993) and
the fast convergence rate (Green 1990) of the penalized EM algorithm,
we present two penalized EM-type algorithms to achieve the PMLE $\tilde{\Psi}$.

Consider the complete data $(X,Z)=\{ X_j,Z_j\}_{j=1}^n$,
where the latent component-indicators vector $Z_{j}=(Z_{1j},\cdots,Z_{pj})$
 follows a multinomial distribution with 1 trial and cell probabilities $\pi_1,\cdots,\pi_p$.
Write it as $Z_{j}\sim \mathcal{M}(1;\pi_1,\cdots,\pi_p)$.
Note that $Z_1,\cdots,Z_n$ are mutually independent.
Based on the component-indicators, for each $X_j(j=1, \cdots, n)$,
a hierarchical representation for skew normal mixtures is given by
\begin{align}\label{hr}
\begin{split}
&X_j|\tau_j,Z_{ij}=1 \sim N\left( \mu_i+\delta(\lambda_i)\tau_j, (1-\delta^2(\lambda_i))\sigma_i^2 \right), \\
&\tau_j|Z_{ij}=1 \sim TN_{[0,+\infty)}\left( 0,\sigma_i^2 \right), \\
&Z_{j}\sim \mathcal{M}(1;\pi_1,\cdots,\pi_p).
\end{split}
\end{align}
where $\delta(\lambda)=\lambda/\sqrt{1+\lambda^2}$ and
$TN_{[0,+\infty)}(0,\sigma^2)$ denotes the truncated normal distribution.
In addition, $\tau_1,\cdots,\tau_n$ are also mutually independent.

According to (\ref{hr}), ignoring additive constants,
the complete data log-likelihood function is
\begin{align}
\begin{split}
\ell_c(\Psi) = \sum_{i=1}^p\sum_{j=1}^n Z_{ij}\bigg\{ &\log(\pi_i)-\log(\sigma_i^2)
-\frac{1}{2}\log(1-\delta^2(\lambda_i)) \\
&-\frac{\tau_j^2-2\delta(\lambda_i)\tau_j(x_j-\mu_i)+(x_j-\mu_i)^2}
{2\sigma_i^2(1-\delta^2(\lambda_i))} \bigg\}.
\end{split}
\end{align}
By Bayesian theorem, we have
$\tau_j|(X_j=x_j,Z_{ij}=1) \sim TN_{[0,+\infty)}(\mu_{\tau_{ij}},\sigma^2_{\tau_i})$,
where
$\mu_{\tau_{ij}}=\delta(\lambda_i)(x_j-\mu_i), \sigma_{\tau_i}=\sigma_i\sqrt{1-\delta^2(\lambda_i)}$.
Thus, for the current parameters $\Psi^{(t)}=(\pi^{(t)}_1,\cdots,\pi^{(t)}_p,\theta^{(t)}_1,\cdots,\theta^{(t)}_p)$
with $\theta^{(t)}_k=(\mu^{(t)}_k,\sigma^{2(t)}_k,\lambda^{(t)}_k)$,
let $\mu^{(t)}_{\tau_{ij}}=\delta(\lambda^{(t)}_i)(x_j-\mu^{(t)}_i)$ and
$\sigma^{(t)}_{\tau_i}=\sigma^{(t)}_i\sqrt{1-\delta^2(\lambda^{(t)}_i)}$.
The ECM algorithm proceeds as follows:

\textbf{E-step}: Compute the conditional expectations
\begin{align*}
&\alpha^{(t)}_{ij}=E\left(Z_{ij}|X_j=x_j,\Psi^{(t)}\right) = \frac{\pi_i^{(t)}f_{SN}(x_j;\theta_i^{(t)})}
{\sum_{k=1}^p \pi_k^{(t)}f_{SN}(x_j;\theta_k^{(t)})}, \\
&\beta^{(t)}_{ij}=E\left(\tau_j|X_j=x_j,Z_{ij}=1,\Psi^{(t)}\right) =
\mu^{(t)}_{\tau_{ij}}+\sigma^{(t)}_{\tau_i}\Delta_{ij}^{(t)},  \\
&\gamma^{(t)}_{ij}=E\left(\tau^2_j|X_j=x_j,Z_{ij}=1,\Psi^{(t)}\right) =
\mu^{2(t)}_{\tau_{ij}}+\sigma^{2(t)}_{\tau_i}+\mu^{(t)}_{\tau_{ij}}\sigma^{(t)}_{\tau_i}
\Delta_{ij}^{(t)}.
\end{align*}
where $\Delta_{ij}^{(t)}=\phi\left(\lambda_i^{(t)}\frac{x_j-\mu^{(t)}_i}{\sigma^{(t)}_i}\right)\bigg/
\Phi\left(\lambda_i^{(t)}\frac{x_j-\mu^{(t)}_i}{\sigma^{(t)}_i}\right)$.
Thus we get $E(Z_{ij}\tau_j|X_j,\Psi^{(t)})=\alpha^{(t)}_{ij}\beta^{(t)}_{ij}$
and $E(Z_{ij}\tau^2_j|X_j,\Psi^{(t)})=\alpha^{(t)}_{ij}\gamma^{(t)}_{ij}$.
Therefore, the objective function can be written as
\begin{align*}
Q(\Psi|\Psi^{(t)}) &= E\left(\ell_c(\Psi)+p_n(\Psi)|X,\Psi^{(t)}\right) \\
&=\sum_{i=1}^p\sum_{j=1}^n \alpha^{(t)}_{ij}\bigg\{ \log(\pi_i)-\log(\sigma_i^2)
-\frac{1}{2}\log(1-\delta^2(\lambda_i)) \\
&\ \ -\frac{\gamma^{(t)}_{ij}-2\delta(\lambda_i)\beta^{(t)}_{ij}(x_j-\mu_i)+(x_j-\mu_i)^2}
{2\sigma_i^2(1-\delta^2(\lambda_i))} \bigg\} + \sum_{k=1}^pp_n(\sigma_k)+\sum_{k=1}^pp_n(\lambda_k).
\end{align*}

\textbf{CM-step} : Maximize $Q(\Psi|\Psi^{(t)})$ with respect to $\Psi$
under the restriction with $\sum_{k=1}^p \pi_k=1$.

\textbf{1}: Update $\pi^{(t)}_i$ by $\pi^{(t+1)}_i=n^{-1}\sum_{j=1}^n \alpha^{(t)}_{ij}$;

\textbf{2}: Update $\mu^{(t)}_i$ by
$$
\mu^{(t+1)}_i = \frac{\sum_{j=1}^n \alpha^{(t)}_{ij}x_j
-\delta(\lambda^{(t)}_i)\sum_{j=1}^n\alpha^{(t)}_{ij}\beta^{(t)}_{ij}}{\sum_{j=1}^n \alpha^{(t)}_{ij}}.
$$

\textbf{3}: Fix $\mu_i=\mu^{(t+1)}_i$, denote $S_{0i}^{(t)}=\sum_{j=1}^n\alpha^{(t)}_{ij}\gamma^{(t)}_{ij}$,
$S_{1i}^{(t)}=\sum_{j=1}^n\alpha^{(t)}_{ij}\beta^{(t)}_{ij}(x_j-\mu^{(t+1)}_i)$
and $S_{2i}^{(t)}=\sum_{j=1}^n\alpha^{(t)}_{ij}(x_j-\mu^{(t+1)}_i)^2$,
with the definition of $p_n(\sigma)$ in (\ref{cop}),
obtain $\sigma^{2(t+1)}_i$ by setting
$$
\sigma^{2(t+1)}_i = \frac{ S_{0i}^{(t)}
-2\delta(\lambda^{(t)}_i)S_{1i}^{(t)}+S_{2i}^{(t)}
+2a_n\left(1-\delta^2(\lambda^{(t)}_i)\right)s_n^2}
{2\left(1-\delta^2(\lambda^{(t)}_i)\right)\left(a_n+\sum_{j=1}^n\alpha^{(t)}_{ij}\right)}
$$

\textbf{4}: Fix $\mu_i=\mu^{(t+1)}_i$ and $\sigma_i=\sigma^{(t+1)}_i$,
with equivalent transformation of
$p_n(\lambda)=-b_n\{\frac{1}{1-\delta^2(\lambda)}+\log(1-\delta^2(\lambda))-1\}$,
and $\lambda^{(t+1)}_i$ is the solution of
$$
-\delta^3(\lambda_i)\sigma^{2(t+1)}_i
\bigg(2b_n+\sum_{j=1}^n \alpha^{(t)}_{ij}\bigg)
+\left(1+\delta^2(\lambda_i)\right)S_{1i}^{(t)}
-\delta(\lambda_i)\bigg(S_{0i}^{(t)}
+S_{2i}^{(t)}-\sigma^{2(t+1)}_i\sum_{j=1}^n \alpha^{(t)}_{ij}\bigg) =0.
$$
\textbf{4}$^*$: For Azzalini's penalty function
$p(\lambda)=-c_1\log(1+c_2\lambda^2)$ where $c_1=0.876,c_2=0.856$ and
$\lambda^{(t+1)}_i$ is obtained by solving
$$
\sigma^{2(t+1)}_i\delta(\lambda_i)\left(1-\delta^2(\lambda_i)\right)
\bigg(\sum_{j=1}^n \alpha^{(t)}_{ij}-\frac{2c_1c_2}{1-(1-c_2)\delta^2(\lambda_i)}\bigg)
+\left(1+\delta^2(\lambda_i)\right)S_{1i}^{(t)}
-\delta(\lambda_i)\bigg(S_{0i}^{(t)}+S_{2i}^{(t)}\bigg) =0.
$$

With some elementary modifications, the ECME algorithm for fitting
the skew normal mixtures can be conducted by replacing
the 4th CM-step with the following CML-step:

\textbf{CML-step}: Calculate
$$
\left(\lambda^{(t+1)}_1,\cdots,\lambda^{(t+1)}_p\right) =
\underset{\lambda_1,\cdots,\lambda_p}{\arg\max} \bigg\{
\sum_{j=1}^n\log\bigg( \sum_{i=1}^p \pi^{(t+1)}_i
f_{SN}\left(x_j;\mu^{(t+1)}_i,\sigma_i^{2(t+1)},\lambda_i\right)
\bigg)+\sum_{i=1}^pp_n(\lambda_i) \bigg\}.
$$

As noted in Lin {\em et al.} (2007b),
the ECME has a faster convergence rate than the ECM
when $p=1$ or $\lambda_1,\cdots,\lambda_p$ are the structure parameters.
But beyond that, the ECM is the better choice for
the one-dimensional search involved in the 4th CM-step as it
is more efficient than the optimization of multi-parameter involved in CML-step.

To monitor convergence, we stop the EM-type algorithm
after the relative change in the objective function is smaller than
a threshold $10^{-6}$.

\textit{Remark}: Compared with the 4th CM-step in Lin {\em et al.} (2007b),
the 4th CM-step of our penalized algorithm shares a similar structure and then the same computational complexity. We also note that the 4$^*$th CM-step
Azzalini's penalty function can significantly reduce the computational complexity.
Moreover, it is worth noting that the penalty
standing for a prior inverse Gamma distribution
also enjoys the advantage of remaining computational efficiency,
see Ciuperca {\em et al.} (2003) and Chen {\em et al.} (2008).
%To tackle the initialization issue, we choose the best run
%in terms of objective function out of $r = 10$ runs with different $K$-means starts.

\section{Simulation Studies}

\subsection{Penalty comparison}

The first numerical simulation is conducted to compare the performance of
our PMLE only with $p_n(\lambda)$ in (\ref{cop}) to that of
the penalized estimator proposed by Azzalini and Arellano-Valle (2013),
who called it MPLE.  For ease of comparison, the parameter values and the sample sizes
are taken to be the same as that in Azzalini and Arellano-Valle (Fig 5. 2013).
That is, $\theta = (0,1,5)$ and $n=\{50, 100, 250, 350, 500, 1000\}$.
Besides, the replication time is $5000$.

\renewcommand{\baselinestretch}{1.0}
\begin{figure}[htb]
\begin{center}
\scalebox{0.9}[0.9]{\includegraphics{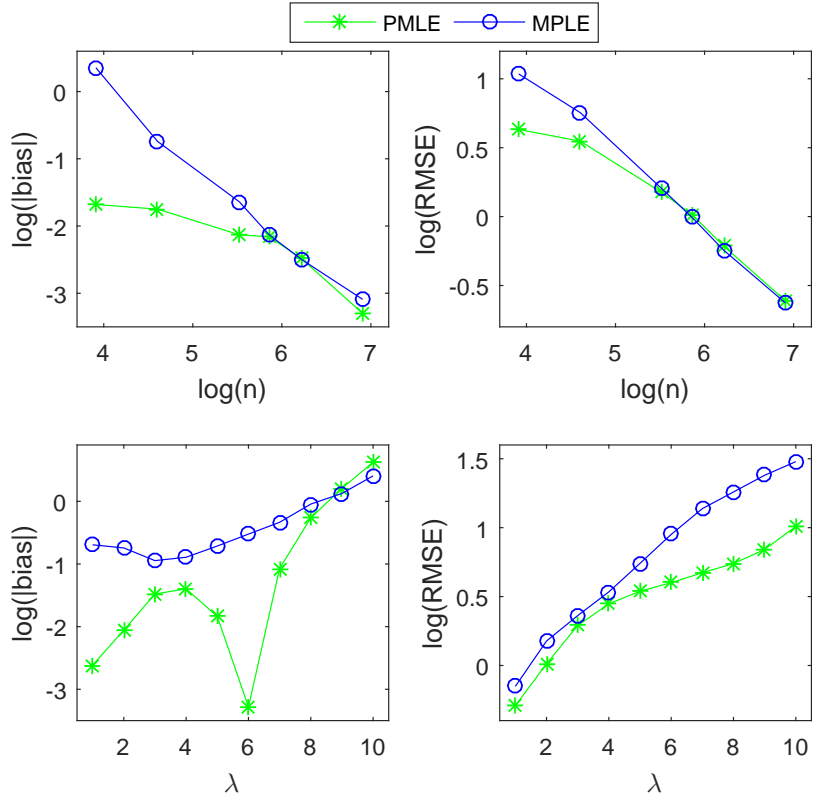}}
\end{center}
\caption{\label{fig:pnt} Simulation study on our PMLE and Azzlini's MPLE.
Top left(right) panel: $\log|\textrm{bias}|$($\log(\textrm{RMSE})$)
are calculated when $\lambda = 5$ and $n=\{50, 100, 250, 350, 500, 1000\}$;
Bottom left(right) panel: $\log|\textrm{bias}|$($\log(\textrm{RMSE})$)
are calculated when $n = 100$ and $\lambda=1,\cdots,10$.}
\end{figure}
\renewcommand{\baselinestretch}{1.5}

The biases and root mean squared  errors (RMSEs) of estimators are
plotted in the two rows of Fig~\ref{fig:pnt}. The first row is with fixed skew value $\lambda=5$  and different sample sizes, while the second row is with the fixed sample size $n=100$ and different $\lambda$. We can then examine how the bias and RMSE can be reduced when the sample size is increasing. This can be showed in the first row of plots indicating the estimation consistency. The second row of plots shows  how they behave when the sample size is fixed to be $n=100$ and the value of $\lambda$ is increasing. PMLE is better performed than MPLE uniformly and the bias of PMLE can be small at some value of $\lambda$ around $6$.
The doubly logarithmic scale is adopted to simplify
the interpretation for the curves.  especially for the upper left penal,
in which the bias of MPLE decreases approximately at the rate of order $n^{-3/2}$.
This is  probably because $p_n(\lambda)$ in (\ref{cop})
 decreases at the rate of order $\log(n)^{-1}$ as $n$ increases,
the bias of PMLE diminishes faster than that of MPLE. Overall, PMLE is markedly preferable to Azzalini's MPLE under small or moderate sample size cases.

%Two bottom panels of Fig \ref{fig:pnt} display the outcomes
%calculated under $n=100$ and $\theta = (0,1,\lambda)$,
%where $\lambda\in\{1,\cdots,10\}$.
%The bottom left panel shows that, when $n=100$,
%PMLE has significant lower bias than MPLE within a wide range,
%except some coincidence at the right end.
%It is also worthy noting that, there is a sharp point
%at $\lambda=6$ in PMLE, where the bias reaches a bottom.
%The behavior reveals an additional benefit of our approach,
%presumably caused by shape of $p_n(\lambda)$.
%In bottom right panel, the obvious improvement of PMLE over MPLE in RMSE
%arises from $\lambda=4$ onwards,
%but the performances are almost identical when $1\leq\lambda\leq4$.
%Thus, the overall indication is that our PMLE is markedly preferable to
%Azzalini's MPLE under small sample size.

\subsection{Simulations for $p=p_0$}

In this subsection, numerical studies are performed to
examine the consistency of the PMLE.
The studies, based on 5000 replications,
consider samples of size $n=\{100,200\}$ from
two 2-component SNMIX models.
The null settings are shown in Table \ref{tbl:para}.

\renewcommand{\baselinestretch}{1.0}
\begin{table}[htb]
\begin{center}
\caption{\label{tbl:para} The settings of two models}
\begin{tabular}{ll}
\toprule
Models  & parameter settings ($SN(\mu,\sigma^2,\lambda)$) \\
\hline
Model I          &$0.5SN(-2,1,2)+0.5SN(2,2,1)$   \\
Model II         &$0.5SN(-1,2,1)+0.5SN(1.5,2,-1)$   \\
\bottomrule
\end{tabular}
\end{center}
\end{table}
\renewcommand{\baselinestretch}{1.5}

For each model, the estimators are obtained by
local maximization of the (penalized) likelihood function
via the (penalized) ECM algorithm. To tackle the initialization issue,
Chen {\em et al.} (2008) used the true mixing distribution
as initial values and Basso {\em et al.} (2010) recommended  a combination of the $K$-means approach and the method of moments.
We employ both schemes in our simulations to see their performance.
In addition, to overcome the effect of label switching (McLachlan \& Peel 2000),
we employ the method on location parameters in Celeux {\em et al.} (1996)
in SNMIX.

\textit{Model I}: The density function of Model I turns out to be bimodal
and well-separated.
Table \ref{tbl:dgn1} shows the minimum of $\hat\sigma^2$,
the maximum of $|\hat\lambda|$ and their degeneracy frequencies
of the two estimators out of 5000 replications.
We regard the estimated values of $\sigma^2$ as 0
when $\hat\sigma^2<10^{-10}$, and take $|\hat\lambda|>100$
as an indication of  divergence.

\renewcommand{\baselinestretch}{1.0}
\begin{table}[htb]
\begin{center}
\caption{\label{tbl:dgn1} Results of parameter estimation for Model I.
(numbers in brackets record the occurrences of $|\hat\lambda|>100$)}
\begin{tabular}{cllcll}
\toprule
 \multirow{2}*{Parameters} &\multicolumn{2}{c}{$n=100$} & &\multicolumn{2}{c}{$n=200$}  \\
 \cline{2-3} \cline{5-6}
 &MLE   &PMLE  &  &MLE   &PMLE   \\
\hline
\multicolumn{6}{l}{\textbf{True values}} \\
$\min(\hat\sigma^2)$       &0.111         &0.136    &   &0.270      &0.273    \\
$\max(|\hat\lambda|)$      &3.8e2(205)    &10.58    &   &1.6e2(4)   &11.03   \\
\multicolumn{6}{l}{\textbf{$K$-means}} \\
$\min(\hat\sigma^2)$       &0.107         &0.133    &   &0.270      &0.272       \\
$\max(|\hat\lambda|)$      &3.7e2(214)    &10.62    &   &1.4e2(3)   &11.02     \\
\bottomrule
\end{tabular}
\end{center}
\end{table}
\renewcommand{\baselinestretch}{1.5}

The outcomes in Table \ref{tbl:dgn1} indicate that
the MLE of $\sigma^2$ does not shrink to $0$ in this case.
However, although the component densities are well-separated,
the MLE still suffers from the divergence on $\lambda$
in both initializations of the algorithms.

\renewcommand{\baselinestretch}{1.0}
\begin{table}[htb]
\begin{center}
\caption{\label{tbl:bsd2} Biases and RMSEs (in brackets) for Model I}
\begin{tabular}{crrcrr}
\toprule
 \multirow{2}*{Parameters} &\multicolumn{2}{c}{$n=100$} & &\multicolumn{2}{c}{$n=200$} \\
 \cline{2-3} \cline{5-6}
&MLE &PMLE & &MLE  &PMLE  \\
\hline
\multicolumn{6}{l}{\textbf{True values}} \\
$\hat\mu_1$        &0.027(0.26)    &0.036(0.24)    &  &0.015(0.19)    &0.020(0.18)    \\
$\hat\mu_2$        &0.071(0.46)    &0.070(0.43)    &  &0.051(0.35)    &0.055(0.35)   \\
$\hat\sigma^2_1$   &0.162(0.74)    &0.081(0.59)    &  &0.072(0.45)    &0.048(0.41)   \\
$\hat\sigma^2_2$   &0.055(0.84)    &0.024(0.77)    &  &0.029(0.62)    &0.017(0.59)   \\
$\hat\lambda_1$    &7.208(40.9)    &0.528(1.82)    &  &0.571(4.36)    &0.299(1.25)   \\
$\hat\lambda_2$    &5.065(32.0)    &0.491(1.62)    &  &0.358(1.94)    &0.233(1.05)   \\
$\hat\pi_1$        &0.008(0.03)    &0.005(0.03)    &  &0.004(0.02)    &0.003(0.02)   \\
\multicolumn{6}{l}{\textbf{$K$-means} } \\
$\hat\mu_1$        &0.088(0.37)    &0.097(0.36)    &  &0.043(0.23)    &0.049(0.22)    \\
$\hat\mu_2$        &0.228(0.79)    &0.226(0.77)    &  &0.117(0.50)    &0.120(0.49)   \\
$\hat\sigma^2_1$   &0.103(0.69)    &0.030(0.56)    &  &0.030(0.45)    &0.006(0.41)   \\
$\hat\sigma^2_2$   &0.293(1.07)    &0.229(0.93)    &  &0.073(0.68)    &0.059(0.65)   \\
$\hat\lambda_1$    &6.385(39.0)    &0.304(1.95)    &  &0.434(4.03)    &0.178(1.29)   \\
$\hat\lambda_2$    &3.522(33.0)    &0.095(2.13)    &  &0.197(1.99)    &0.091(1.25)   \\
$\hat\pi_1$        &0.003(0.03)    &0.001(0.03)    &  &0.002(0.02)    &0.001(0.02)   \\
\bottomrule
\end{tabular}
\end{center}
\end{table}
\renewcommand{\baselinestretch}{1.5}

Table \ref{tbl:bsd2} shows the biases and RMSEs of the two estimators.
It is clear that all biases and RMSEs
of the PMLE in Table~\ref{tbl:bsd2} decrease as $n$ increases,
reflecting its consistency.
It is also remarkable that, the PMLE is far superior
in the performances of estimating $\lambda_1 $ and $\lambda_2$ to the MLE,
remedy the indeed unreliable MLE of $\lambda$
especially when $n$ is small. Overall, the PMLE significantly outperforms the MLE except for the mean $\mu_1$.
Meanwhile, presumably due to the well separate kernel densities,
the MLEs and PMLEs of all other parameters work well.

\renewcommand{\baselinestretch}{1.0}
\begin{figure}[th]
\begin{center}
\scalebox{0.7}[0.7]{\includegraphics{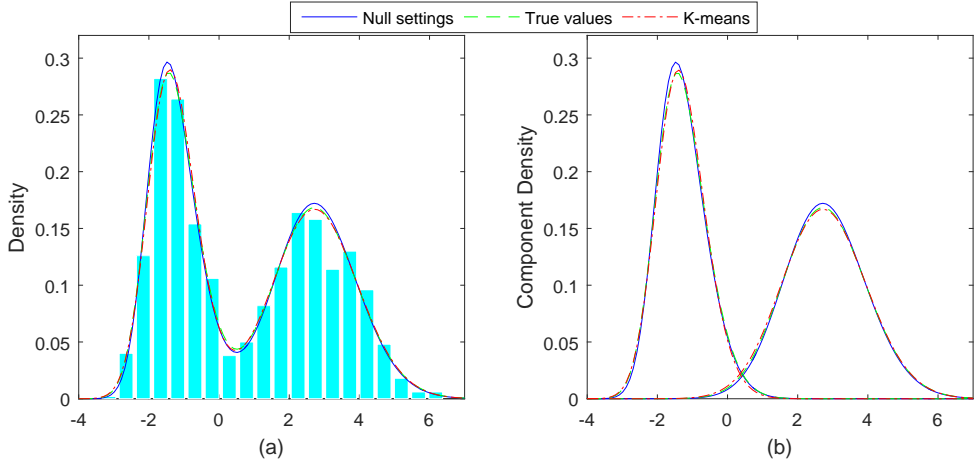}}
\end{center}
\caption{\label{fig:1000} (a) Histogram of the simulating data from Model I
with $n=1000$ overlaid with densities under null settings and
MLEs with two starts,
(b) Component densities under null settings and MLEs with two starts}
\end{figure}
\renewcommand{\baselinestretch}{1.5}

The unreliable MLE of $\lambda$ reaffirms the theoretical expectation
in DiCiccio and Monti (2004). That is, larger sample sizes
would be required to improve the estimation accuracy.
When $n=1000$, the mixing distribution can generally be
reliably estimated by the MLE.
A separate simulation study is performed in Model I
with $n=1000$, the outcome is summarized graphically in Fig \ref{fig:1000}.
In this situation, not only the estimated distribution $\hat\Psi$
but also all elements in $\hat\Psi$ converge to that of $\Psi_0$.

Moreover, for SN distribution, DiCiccio and Monti (2004)
proposed an easy-to-implement procedure to handle the estimation divergence about  $\lambda$.
They defined $\hat\lambda$  by the smallest value
$\breve\lambda$ such that $H_0:\lambda=\breve\lambda$
is not rejected at the $5\%$ nominal level by a profile likelihood ratio test
converging in distribution to $\chi^2_1$.

Let $\hat\Lambda$ be the MLE of $\Lambda=(\lambda_1,\cdots,\lambda_p)$
with $\Lambda_0=(\lambda_{01},\cdots,\lambda_{0p})$.
Noting that there is no $\sigma^2\rightarrow0$ in Model I,
we thus extend the approach of DiCiccio and Monti (2004) to
SNMIX by taking $\hat\Lambda$ to be a modified estimator (ME)
$\breve\Lambda=(\breve\lambda_1,\cdots,\breve\lambda_p)$.
The ME $\breve\Lambda$ is obtained by
$\max_{\breve\Lambda}||\breve\Lambda-\hat\Lambda||_1$
under that the composite null hypothesis $H_0: \Lambda=\breve\Lambda$
is not rejected at the $5\%$ nominal level by
the profile likelihood ratio test whose limiting null distribution is
chi-squared $\chi^2_\nu$ with the degrees of freedom $\nu=\sum_{k=1}^{p}I(|\lambda_k|\geq30)$.

\renewcommand{\baselinestretch}{1.0}
\begin{table}[htb]
\begin{center}
\caption{\label{tbl:me} Biases and RMSEs (in brackets) of the ME for Model I on $\lambda$.}
\begin{tabular}{cllcll}
\toprule
 \multirow{2}*{Parameters} &\multicolumn{2}{c}{True values} &
 &\multicolumn{2}{c}{$K$-means}  \\
 \cline{2-3} \cline{5-6}
 &$n=100$   &$n=200$  &  &$n=100$   &$n=200$   \\
\hline
$\breve\lambda_1$      &1.247(3.87)    &0.443(1.74)      &   &0.972(3.91)    &0.317(1.76)      \\
$\breve\lambda_2$      &0.967(3.19)    &0.336(1.52)      &   &0.446(3.75)    &0.183(1.64)     \\
\bottomrule
\end{tabular}
\end{center}
\end{table}
\renewcommand{\baselinestretch}{1.5}

The only difference between the ME and MLE is mainly on the performance of
estimating the shape parameters $\lambda_1$ and $\lambda_2$.
Table~\ref{tbl:me} shows that  with both the initial schemes of the algorithms,
the ME has power to exclude  diverging estimates.
However, as shown in Table~\ref{tbl:bsd2}, the PMLE  works better than the ME.

\textit{Remark}: Unfortunately, the extension of DiCiccio and Monti's (2004) approach
to the SNMIX lacks rigorous theoretical basis.
When $\breve\sigma^2=\hat\sigma^2\rightarrow0$,
the null hypothesis $H_0: \Lambda=\breve\Lambda$ lies on
the boundary of the parameter space. That is,
the regularity conditions are not satisfied for the mixture problem
considered here, and the asymptotic $\chi^2$ theory of the
likelihood ratio test statistic does not hold.
Hence, their method remains invalid in SNMIX but
it seems to be applicable only in the well-separated case without $\hat\sigma^2\rightarrow0$,
displaying an additional advantage of the penalized estimator.

In the following, we consider a more difficult situation
in which two component densities in the mixture are close to one another. However, an interesting observation is that although the parameters cannot be well estimated separately, the densities can be estimated accurately.

\textit{Model II}: The components of Model II are of homoscedasticity
and the density function here seems to be strongly
unimodal and thus poorly-separated.
The simulated results are presented in Tables \ref{tbl:dgn} and \ref{tbl:bsd1}.

\renewcommand{\baselinestretch}{1.0}
\begin{table}[htb]
\begin{center}
\caption{\label{tbl:dgn} Results of parameter estimation for Model II (the
numbers in brackets record the occurrences of $\hat\sigma^2<$1e-10$(1\times10^{-10})$
and $|\hat\lambda|>100$ respectively)}
\begin{tabular}{cllcll}
\toprule
 \multirow{2}*{Parameters} &\multicolumn{2}{c}{$n=100$} & &\multicolumn{2}{c}{$n=200$}  \\
 \cline{2-3} \cline{5-6}
 &MLE  &PMLE  &  &MLE  &PMLE   \\
\hline
\multicolumn{6}{l}{\textbf{True values}} \\
$\min(\hat\sigma^2)$       &7.8e-31(62)   &0.008     &  &7.7e-31(12)   &0.004    \\
$\max(|\hat\lambda|)$      &5.3e2(533)    &9.511     &  &3.3e2(91)     &11.01   \\
\multicolumn{6}{l}{\textbf{$K$-means}} \\
$\min(\hat\sigma^2)$       &2.7e-304(3)   &0.008     &  &0.001         &0.009       \\
$\max(|\hat\lambda|)$      &3.1e2(638)    &11.14     &  &2.9e2(75)     &12.28     \\
\bottomrule
\end{tabular}
\end{center}
\end{table}
\renewcommand{\baselinestretch}{1.5}

It can be observed from Table \ref{tbl:dgn} that
(a) the MLE suffers from degeneracies in both of $\sigma^2$ and $\lambda$
even when the true distribution is used for initialization of the algorithm,
the difficult situation is eased as sample size increases;
(b) the penalized approach solves both degenerate problems on
$\sigma^2$ and $\lambda$, all estimated values of PMLE
are well confined; (c) the classical clustering procedure $K$-means
has, to certain extent, an excluding effect in fitting $\hat\Psi$
with degenerate component variances as we found that the $K$-means initialization
of the algorithm can reduce the proportion of diverging values compared
with the true value initialization.

Table \ref{tbl:dgn} also indicates a remarkable higher
degenerate frequency on $\hat\lambda$  than on $\hat\sigma^2$.
The divergence of shape parameters in SNMIX thus must be
paid more attention in practice.
The phenomenon can also partly explain our use of a significantly slower decreasing $p_n(\lambda)$
defined in (\ref{tp}) with rate $(\log n)^{-1}$,
than the $n^{-1}$ of $p_n(\sigma)$ as $n$ increases.

\renewcommand{\baselinestretch}{1.0}
\begin{table}[h]
\begin{center}
\caption{\label{tbl:bsd1} Biases and RMSEs (in brackets) for Model II}
\begin{tabular}{crrcrr}
\toprule
 \multirow{2}*{Parameters} &\multicolumn{2}{c}{$n=100$} & &\multicolumn{2}{c}{$n=200$} \\
 \cline{2-3} \cline{5-6}
&MLE &PMLE & &MLE  &PMLE  \\
\hline
\multicolumn{6}{l}{\textbf{True values}} \\
$\hat\mu_1$        &-0.123(0.74)   &-0.094(0.72)   &  &-0.081(0.57)   &-0.063(0.54)    \\
$\hat\mu_2$        &0.154(0.76)    &0.125(0.73)    &  &0.089(0.58)    &0.075(0.56)   \\
$\hat\sigma^2_1$   &-0.901(5.88)   &-0.579(1.09)   &  &-0.279(2.28)   &-0.280(0.72)   \\
$\hat\sigma^2_2$   &-0.793(5.15)   &-0.595(1.12)   &  &-0.319(2.63)   &-0.288(0.74)   \\
$\hat\lambda_1$    &12.14(52.1)    &0.303(1.32)    &  &2.707(19.8)    &0.286(1.19)   \\
$\hat\lambda_2$    &-13.02(54.4)   &-0.350(1.36)   &  &-2.497(18.5)   &-0.269(1.13)   \\
$\hat\pi_1$        &0.003(0.23)    &0.005(0.25)    &  &0.003(0.19)    &0.004(0.19)   \\
\multicolumn{6}{l}{\textbf{$K$-means} } \\
$\hat\mu_1$        &0.871(1.18)    &0.846(1.20)    &  &0.947(1.13)    &0.935(1.14)    \\
$\hat\mu_2$        &-0.808(1.14)   &-0.784(1.17)   &  &-0.951(1.13)   &-0.940(1.13)   \\
$\hat\sigma^2_1$   &-0.719(1.52)   &-0.745(1.13)   &  &-0.562(0.87)   &-0.582(0.89)   \\
$\hat\sigma^2_2$   &-0.920(10.0)   &-0.796(1.19)   &  &-0.543(0.82)   &-0.564(0.83)   \\
$\hat\lambda_1$    &-15.27(50.5)   &-2.752(3.57)   &  &-4.648(15.7)   &-2.414(3.29)   \\
$\hat\lambda_2$    &14.64(50.1)    &2.699(3.52)    &  &4.447(16.8)    &2.411(3.29)   \\
$\hat\pi_1$        &0.010(0.24)    &0.009(0.25)    &  &0.001(0.20)    &0.001(0.20)   \\
\bottomrule
\end{tabular}
\end{center}
\end{table}
\renewcommand{\baselinestretch}{1.5}

Table \ref{tbl:bsd1} reports the biases and RMSEs of
the MLE and PMLE under Model~II.
To manifest the discrepancy between degenerate $\hat\sigma^2$
and $\sigma^2_0$, and to  make a sensible comparison,
we calculate the bias and RMSE of $\log(\hat\sigma^2_i)$
instead of $\hat\sigma^2_i$, which are in proportion to
the relative indicators used in Chen {\em et al.} (2008).
When $\Psi_0$ is used for initialization,
the biases and RMSEs of PMLE reduce rapidly as $n$ increases,
with remarkable superiority of PMLE over MLE displayed
on $\hat\lambda$ and $\hat\sigma^2$.

In the case of $K$-means initialization, although $\hat \sigma^2\rightarrow0$
has been largely prevented, $\Psi$ has not been estimated accurately,
both the estimators even lose the consistency on $\hat\mu$.
This behavior was investigated through a separate simulation study,
which is conducted on a data set generated from Model II with $n=1000$.
Table~\ref{tbl:tk1000} shows the MLE and PMLE are almost equivalent
under the same initialization scheme,
while quite different if the initialization changes.
Meanwhile, the values of $p\ell_n(\hat\Psi)$ obtained
when $\Psi_0$ is used as initial value are smaller than
those that are based on  $K$-means initialization. That is, the EM-type algorithm converges to
a local maximum when starting from $\Psi_0$,
the $K$-means based estimates seem the global maximum solution,
which leads to the poor performances in Table~\ref{tbl:bsd1}.

\begin{figure}[th]
\begin{center}
\scalebox{0.7}[0.7]{\includegraphics{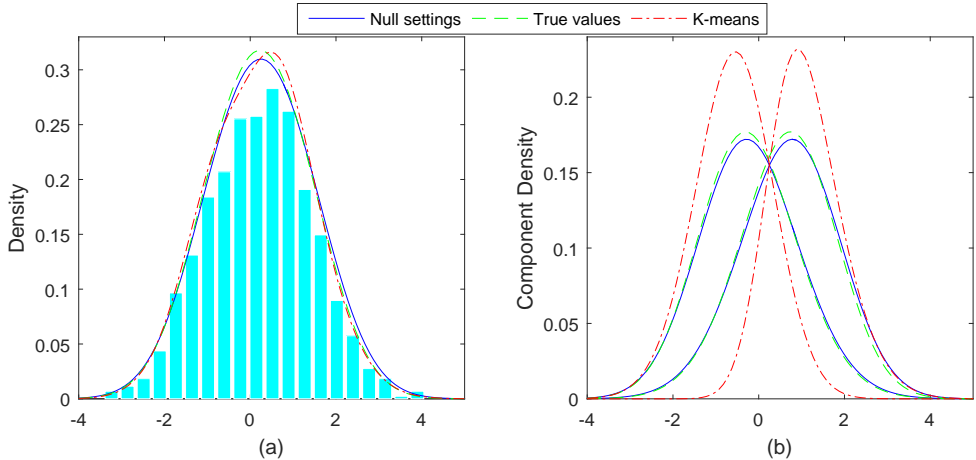}}
\end{center}
\caption{\label{fig:tk1000} (a) Histogram of the simulating data from Model II
with $n=1000$ overlaid with densities under null settings and MLEs with two starts,
(b) Component densities under null settings and MLEs with two starts}
\end{figure}

The outcomes in Table~\ref{tbl:tk1000} are vividly
summarized in Figure~\ref{fig:tk1000}.
We can see that although the $K$-means based fitted mixing density is close to the true value based estimate,
the resulting component densities differ substantially
from the true ones. In other words, this phenomenon does not challenge the identifiability of
finite mixture models(Wald 1949; Kiefer and Wolfowitz 1956),
but reveals  the so-called "\textit{over-flexibility}" shortcoming of the estimation methods for SNMIX
when the two mixing components are close to one another.
%Thus in practice, the skew normal mixtures should be used with caution.

\renewcommand{\baselinestretch}{1.0}
\begin{table}[htb]
\begin{center}
\caption{\label{tbl:tk1000} Parameter estimates for Model II when $n=1000$}
\begin{tabular}{ccccccccc}
\toprule
Method &$\mu_1$  &$\mu_2$  &$\sigma^2_1$  &$\sigma^2_2$  &$\lambda_1$  &$\lambda_2$
&$\pi_1$  &$p\ell_n(\hat\Psi)$  \\
\hline
\multicolumn{9}{l}{\textbf{True values}} \\
MLE      &-1.018   &1.440   &1.940   &1.873   &1.018   &-1.016   &0.504   &-1617   \\
PMLE     &-1.018   &1.439   &1.940   &1.873   &1.018   &-1.015   &0.504   &-1617   \\
\multicolumn{9}{l}{\textbf{$K$-means} } \\
MLE      &-0.037   &0.273   &1.136   &1.399   &-0.846  &1.814    &0.525   &-1616   \\
PMLE     &-0.038   &0.273   &1.136   &1.398   &-0.844  &1.804    &0.525   &-1616    \\
\bottomrule
\end{tabular}
\end{center}
\end{table}
\renewcommand{\baselinestretch}{1.5}

\renewcommand{\baselinestretch}{1.0}
\renewcommand{\baselinestretch}{1.5}

Another interesting observation about the $K$-means based estimation
is its much lower occurance of $\hat\sigma^2\rightarrow0$ than the true value
based estimation in SNMIX presented in Table~\ref{tbl:dgn} in spite of
its poor performances indicated in Table~\ref{tbl:bsd1}.
To further study this phenomenon,
we consider  Example 2 of Chen {\em et al.} (2008),
in which data were generated from $0.5N(0,1)+0.5N(1.5,3)$.
In this case, based on the two starting strategies,
we fit the data with both two-component GMIX and SNMIX.
The replication time is again $5000$.

\renewcommand{\baselinestretch}{1.0}
\begin{table}[htb]
\begin{center}
\caption{\label{tbl:dgnc} Results of parameter estimation for $0.5N(0,1)+0.5N(1.5,3)$
fitted by GMIX and SNMIX. (the numbers in brackets record the occurrences
of $\hat\sigma^2<$1e-10 and $|\hat\lambda|>100$ respectively)}
\begin{tabular}{cllcll}
\toprule
 \multirow{2}*{Parameters} &\multicolumn{2}{c}{True values} & &\multicolumn{2}{c}{$K$-means}  \\
 \cline{2-3} \cline{5-6}
 &$n=100$  &$n=200$  &  &$n=100$  &$n=200$   \\
\hline
\multicolumn{6}{l}{\textbf{GMIX}} \\
$\min(\hat\sigma^2)$       &0(30)   &0(3)   &  &0(40)  &0(3)    \\
\multicolumn{6}{l}{\textbf{SNMIX}} \\
$\min(\hat\sigma^2)$    &4.3e-252(22)  &7.7e-31(2)   &  &7.8e-304(3)  &3.1e-30(2)  \\
$\max(|\hat\lambda|)$   &3.0           &2.8          &  &4.4e2(588)   &3.0e2(116)     \\
\bottomrule
\end{tabular}
\end{center}
\end{table}
\renewcommand{\baselinestretch}{1.5}

The results in Table \ref{tbl:dgnc} suggest the following.
For the GMIX model, the $K$-means based estimation gets more degenerate $\hat\sigma^2$
than the true value based estimation does when $n=100$.
While for  SNMIX, the $K$-means based estimation avoids $\hat\sigma^2\rightarrow0$
more efficiently than the true value based estimation.
However, the $K$-means based estimation suffers much more severely from
the divergence of $\hat\lambda$ than the true value based estimation.
This may explain the reason why the aggregated estimation effect of
the $K$-means based estimation is worse.

% This example does show that under the SNMIX structure, compared to GMIX, K-means based estimation has lower  occurance of $\hat\sigma^2\rightarrow0$ than the true value based estimation.

The achievement of preventing $\hat\sigma^2\rightarrow0$
under Model~II thus comes from the cooperation of the SNMIX modelling scheme
and the $K$-means starting strategy.
An intuitive explanation for this phenomenon is that,
the $K$-means starts in SNMIX can escape from
the attraction domain  around the singularities,
the existence of which was proved by Biernacki and Chr¨¦tien (2003).

\subsection{Simulation for $p>p_0$}

In the case of $p_0<p<\infty$, for convenience,
the data have still been sampled from Model I with $n=\{100, 200, 500\}$,
that is, $p_0=2$. For each data set, the MLE and PMLE are computed
when $p=\{2, 3, 4, 5\}$. The simulating size is $1000$.

Since $p> p_0$, we cannot expect that
every part of $\hat\Psi$ equals to that of $\Psi_0$.
To handle this situation, Chen {\em et al.} (2008) and Chen \& Tan (2009)
employed ten values in the neighbourhood of $\Psi_0$
as the starts of the ECM algorithm.
In this simulation, the ten initial values are obtained by
slightly perturbing $\mu_{0j}$ in $\Psi_0$.
The perturbation proceeds as follows:
$$
\mu_{i}=\mu_{0j}+N(0,0.1^2),\pi_{i}=\pi_{0j}/\omega_j;i=1,\cdots,p;j=1,\cdots,p_0.
$$
where $\omega_j$ is the total component number that $\mu_{i}$ comes from $\mu_{0j}$
and $\sum_{j=1}^{p_0}\omega_j=p$.
Given the initial values, the best run in terms of objective function
is taken as the final estimator.

In this case, it is meaningful to investigate the distance $D(\hat\Psi,\Psi_0)$
defined in (\ref{dst}). However, it is not sensible for measuring the discrepancy
between $\hat\Psi$ and $\Psi_0$.
To improve the situation, we employ a modified distance
$D^*(\hat\Psi,\Psi_0)=\int_{\Theta^*}|\hat\Psi(\theta)-\Psi_0(\theta)|d\theta$,
where $\theta = (\mu,\log(\sigma^2)/5,\log(\lambda)/2)$ and
$\Theta^*=[-5,10]\times[-15,1]\times[-10,5]$.
Note that all parameter values of two estimators
are within the region $\Theta^*$.

The numbers of degeneracies of MLE are shown in Table \ref{tbl:nd}.
It is immediately clear that the frequencies of degeneracy
of $\sigma^2$ and $\lambda$ decrease as $n$ increases and
increase as the putative order $p$ increases.
In addition, we also observe a higher frequency of degeneracy
existing on $\lambda$ over $\sigma^2$,
in agreement with Table \ref{tbl:dgn1} and \ref{tbl:dgn}.
We also issue a statement here that, in all cases,
there is no degenerate outcomes occurred in our penalized estimator.

\renewcommand{\baselinestretch}{1.0}
\begin{table}[htb]
\begin{center}
\caption{\label{tbl:nd}Number of degeneracies in $\hat\sigma^2$ and $\hat\lambda$ for Model I.}
\begin{tabular}{cllcllcll}
\toprule
 \multirow{2}*{$p_0=2$} &\multicolumn{2}{c}{$n=100$}  &  &\multicolumn{2}{c}{$n=200$}
 &  &\multicolumn{2}{c}{$n=500$}  \\
 \cline{2-3}  \cline{5-6}  \cline{8-9}
   &$\hat\sigma^2$  &$\hat\lambda$  &  &$\hat\sigma^2$  &$\hat\lambda$ &
   &$\hat\sigma^2$  &$\hat\lambda$  \\
\hline
$p=2$     &0     &40     &   &0    &2      &   &0   &0     \\
$p=3$     &12    &492    &   &3    &134    &   &0   &1     \\
$p=4$     &79    &901    &   &12   &290    &   &0   &12    \\
$p=5$     &166   &1196   &   &30   &437    &   &0   &25    \\
\bottomrule
\end{tabular}
\end{center}
\end{table}
\renewcommand{\baselinestretch}{1.5}

\renewcommand{\baselinestretch}{1.0}
\begin{table}[htb]
\begin{center}
\caption{\label{tbl:dst}Average $D^*(\Psi,\Psi_0)$ of MLE and PMLE for Model I}
\begin{tabular}{crrrcrrr}
\toprule
\multirow{2}*{$p_0=2$} &\multicolumn{3}{c}{MLE} &  &\multicolumn{3}{c}{PMLE}  \\
 \cline{2-4}  \cline{6-8}
  &$n=100$  &$n=200$  &$n=500$  &  &$n=100$  &$n=200$  &$n=500$  \\
\hline
$p=2$     &9.46    &6.84     &4.35    &   &8.17    &6.53     &4.32    \\
$p=3$     &15.56   &10.83    &6.59    &   &10.37   &9.00     &6.23    \\
$p=4$     &19.04   &13.39    &7.95    &   &12.29   &10.83    &7.48    \\
$p=5$     &22.28   &15.44    &9.08    &   &13.82   &12.19    &8.48    \\
\bottomrule
\end{tabular}
\end{center}
\end{table}
\renewcommand{\baselinestretch}{1.5}

Table \ref{tbl:dst} reports the averages of $D^*(\hat\Psi,\Psi_0)$.
In each case, the mean of $D^*(\hat\Psi,\Psi_0)$ decreases as $n$ increases.
The slow decreasing rate of $D^*(\hat\Psi,\Psi_0)$ may be also explained
by the conclusion in Chen (1995) that the optimal convergence rate of
estimated distribution is at most $n^{-1/4}$ when $p>p_0$.
Moreover, for $n=100$, we can observe a significantly smaller
and slower increasing average distance of PMLE over MLE,
indicates the superiority of PMLE when $p>p_0$.
However, the discrepancies on average $D^*(\hat\Psi,\Psi_0)$ of two methods
gradually vanishes as $n$ increases.

\section{Application Examples}

\subsection{Body mass index data}

This data set contains information about body mass index (BMI),
an important medical standard used to measure obesity,
calculated by the ratio of body weight ($\textrm{kg}$)
and square of body height ($\textrm{m}^2$).
The BMI data is collected by the National Health and Nutrition
Examination Survey, conducted annually by the National Center
for Health Statistics of the Center for Disease Control in the USA.
According to the reports in years 1999-2000 and 2001-2002,
Lin {\em et al.} (2007a) investigated the BMI of man participants aged
between 18 to 80, whose weights are lying within
$[39.5, 70]\ \textrm{kg}$ and $[95.01, 196.8]\ \textrm{kg}$.
The data is strongly bimodal and thus fitted with two-component mixtures
using four distributions: Normal, Student' $t$, Skew normal and Skew $t$.
Another two distributions, skew contaminated normal and skew slash distribution,
are introduced by Prates {\em et al.} (2013) to model this data.

We obtain the BMI data consisting of 2107 participants in
R package \textbf{mixsmsn} presented by Prates {\em et al.} (2013).
To compare the proposed PMLE with the ordinary MLE,
the parameter estimations and penalized log-likelihoods are displayed
in Table \ref{tbl:bmi}. The results are the best performer of objective functions
out of 20 runs with different $K$-means starts. A relative tolerance of
$10^{-6}$ for objective functions is employed in the ECM algorithm as
the convergence criterion.

\renewcommand{\baselinestretch}{1.0}
\begin{table}[htb]
\begin{center}
\caption{\label{tbl:bmi} Parameter estimates for BMI data}
\begin{tabular}{lcccccccc}
\toprule
Method &$\mu_1$  &$\mu_2$  &$\sigma^2_1$  &$\sigma^2_2$  &$\lambda_1$  &$\lambda_2$
&$\pi_1$  &$p\ell_n(\hat\Psi)$  \\
\hline
MLE      &19.70   &28.71   &12.45  &62.80  &1.622   &8.104   &0.522   &-6870   \\
PMLE     &19.74   &28.70   &12.05  &62.69  &1.564   &7.618   &0.520   &-6870   \\
\bottomrule
\end{tabular}
\end{center}
\end{table}
\renewcommand{\baselinestretch}{1.5}

The results of MLE and PMLE are given in Table \ref{tbl:bmi},
and they are essentially equivalent to each other.
Thus in the case without degeneracies on $\sigma^2$ and $\lambda$,
the PMLE can be sufficiently close to MLE.
Besides, it appears that the fitted model is of
significant heteroscedasticity.
The approximation of two approaches in this data set seems reasonable,
since the effect of the penalizing terms $p_n(\sigma)$ and $p_n(\lambda)$
naturally disappear as $n$ increases to infinity.

\subsection{The Faithful data}

For the second case, we investigate the accuracy of the proposed PMLE
in a data set with small sample size. A good choice is the famous Faithful data,
which is collected from Old Faithful Geyser in Yellowstone National Park.
Scientists presented analysis on this data,
see Silverman (1986) and Azzalini and Bowman (1990).
It consists of 272 observations, measured on two variables (in minutes):
eruption length and eruption duration.
Lin {\em et al.} (2007b) and Prates {\em et al.} (2013)
fitted the data with univariate and bivariate two-component SNMIX respectively,
both of which have better performance than corresponding two-component GMIX.

\begin{figure}[th]
\begin{center}
\scalebox{0.8}[0.8]{\includegraphics{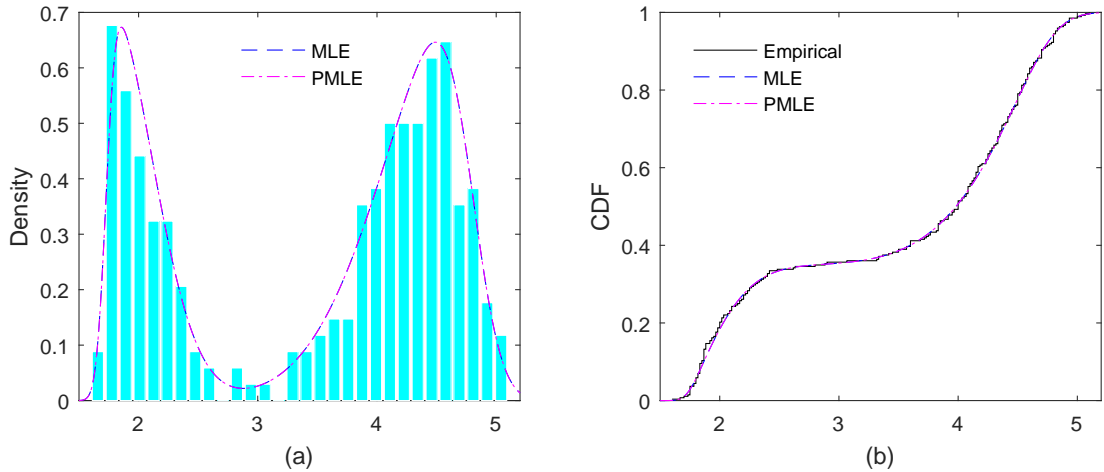}}
\end{center}
\caption{\label{fig:fath} (a) Histogram of the faithful data overlaid
with densities based on the results of MLE and PMLE,
(b) Empirical CDF of the faithful data overlaid
with CDFs based on MLE and PMLE }
\end{figure}

We focus on fitting eruption length with two-component SNMIX
and list the outcomes in the Table \ref{tbl:fath}.
As expected, all parameter values of PMLE
keep almost the same as that of the MLE.
The similarity of MLE and PMLE is also reemphasized by
the density curves and CDF curves in Figure \ref{fig:fath},
both of which are inseparable.

\renewcommand{\baselinestretch}{1.0}
\begin{table}[htb]
\begin{center}
\caption{\label{tbl:fath} Parameter estimates for Faithful data}
\begin{tabular}{lcccccccc}
\toprule
Method  &$\mu_1$  &$\mu_2$  &$\sigma^2_1$   &$\sigma^2_2$
&$\lambda_1$  &$\lambda_2$  &$\pi_1$  &$p\ell_n(\hat\Psi)$ \\
\hline
MLE       &1.727   &4.796  &0.145  &0.463  &5.818  &-3.401   &0.349   &-257.9   \\
PMLE      &1.728   &4.794  &0.143  &0.462  &5.559  &-3.357   &0.349   &-257.9   \\
\bottomrule
\end{tabular}
\end{center}
\end{table}
\renewcommand{\baselinestretch}{1.5}

\renewcommand{\baselinestretch}{1.0}
\renewcommand{\baselinestretch}{1.5}

Based on the results of Table \ref{tbl:fath} and Figure \ref{fig:fath},
it strongly suggests that the proposed penalizing approach can give
a reasonably accurate estimate, which would be also sufficient for data sets
of small sample size.

\section{Conclusions}

In this paper, we propose a penalized MLE to overcome both the degeneracy of $\sigma^2$
and the divergence of $|\lambda|$ in MLE in skew noraml mixture models.
The rigorous proofs of the consistency of the PMLE are provided
when the putative order $p$ is equal to or larger than $p_0$.
The approach developed could be valid for regaining the consistency
and efficiency, and have the advantage of placing no additional constraint
on the parameter space.
This methodology can be extensively applicable to other class of finite mixture
models, for example, the multivariate SNMIX models (Lin 2009) and finite mixture
of skew-$t$ distributions (Lin {\em et al.} 2007a),
which are more complicated and worthy of further investigation.

\section*{Appendix}
\renewcommand{\theequation}{A.\arabic{equation}}

\subsection*{Proof of Theorem \ref{thm1}}
\begin{proof}
Define index sets of observations
$A_{(k)}=\{ i:|x_i-\mu_{(k)}|<|\sigma_{(k)}\log\sigma_{(k)}| \}$
for $k=1,\cdots,p$.
For any set $S$, let $n(S)$ be the number of elements in $S$ and define
$\ell_n(\Psi;S) = \sum_{i\in S} \log f(x_i;\Psi)$.

For $\Psi\in\Gamma_{\sigma}^p$ and small enough $\epsilon_0$,
the mixture density $ f(x_i;\Psi)\leq\frac{1}{\sigma_{(k)}}$ for any $i\in A_{(k)}$.
Since $n(\cap_{t=1}^{k-1}A_{(t)}^c\cap A_{(k)})\leq n(A_{(k)})$,
recall the bound for $n(A_{(k)})$ in lemma \ref{nn2},
almost surely, we have
\begin{equation}\label{ll}
\ell_n(\Psi;\cap_{t=1}^{k-1}A_{(t)}^c\cap A_{(k)}) \leq -n(A_{(k)})\log\sigma_{(k)}
\leq 4Mn\sigma_{(k)}\log^2\sigma_{(k)}-10\log\sigma_{(k)}\log n,
\end{equation}

Adding penalty function $p_n(\sigma_{(k)})$ satisfying conditions
$\mathbf{C1}$-$\mathbf{C2}$, the (\ref{ll}) can be extended as
\begin{align}\label{lp}
\begin{split}
\ell_n & (\Psi;\cap_{t=1}^{k-1}A_{(t)}^c\cap A_{(k)})+p_n(\sigma_{(k)}) \\
&\leq 4Mn\sigma_{(k)}\log^2\sigma_{(k)}-(10\log n-\log^2 n)\log\sigma_{(k)} \\
&\leq 4Mn\sigma_{(k)}\log^2\sigma_{(k)} \leq 4Mn\epsilon_0\log^2\epsilon_0.
\end{split}
\end{align}

For any $i\in\cap_{t=1}^{p}A_{(t)}^c$,
since $|x_i-\mu_{(k)}|>|\sigma_{(k)}\log\sigma_{(k)}|$,
it is easy to show
$$
\log f(x_i;\Psi)
\leq \log\bigg\{ \sum_{k=1}^p \frac{2\pi_k}{\sigma_k}\phi\bigg(\frac{x_i-\mu_k}{\sigma_k}\bigg) \bigg\}
\leq \log\bigg\{ \sum_{k=1}^p \frac{2\pi_k}{\sigma_k}\phi(-\log\sigma_k) \bigg\}
\leq -\log\epsilon_0-\frac{\log^2\epsilon_0}{2} <0
$$
By $4pM\epsilon_0\log^2\epsilon_0\leq1$,
$-4pM\epsilon_0\log\epsilon_0\leq\frac{p-1}{p}$
holds for small enough $\epsilon_0$, this further implies
$$
n(\cap_{t=1}^{p}A_{(t)}^c)\geq n-\sum_{t=1}^{p}n(A_{(t)})\geq \frac{n}{p}.
$$
Hence, the total log-likelihood contributions of observations
in $\cap_{t=1}^{p}A_{(t)}^c$ are bounded by
\begin{equation}\label{lp1}
\ell_n(\Psi;\cap_{t=1}^{p}A_{(t)}^c)
\leq -\frac{n}{p}\bigg\{\log\epsilon_0+\frac{(\log\epsilon_0)^2}{2}\bigg\}
\end{equation}

Thus, for $\Psi\in\Gamma_{\sigma}^p$ and the selected sufficiently small $\epsilon_0$,
with the results of (\ref{lp}) \& (\ref{lp1}) and condition $\mathbf{C3}$,
the penalized log-likelihood has the upper bound as
\begin{align*}
p\ell_n(\Psi) &= \sum_{k=1}^p \left\{\ell_n(\Psi;\cap_{t=1}^{k-1}A_{(t)}^c\cap A_{(k)})+p_n(\sigma_{(k)})\right\}
+\ell_n(\Psi;\cap_{t=1}^{p}A_{(t)}^c) +o(n)\\
&\leq 4pMn\epsilon_0\log^2\epsilon_0-\frac{n}{p}\bigg\{\log\epsilon_0+\frac{(\log\epsilon_0)^2}{2}\bigg\}+o(n) \\
&\leq n+n(K_0-2)+o(n)= n(K_0-1)+o(n)
\end{align*}

By the strong law of large numbers,
we have $\frac{1}{n}p\ell_n(\Psi_0)\underset{\longrightarrow}{a.s.} K_0$.
Consequently, as $n\rightarrow\infty$, almost surely,
$$
\sup_{\Gamma_{\sigma}^p} p\ell_n(\Psi)-p\ell_n(\Psi_0) \leq -n+o(n)\rightarrow -\infty .
$$

\end{proof}

\subsection*{Proof of Theorem \ref{thm2}}
\begin{proof}
Let $\bar{\Gamma}_{\sigma}^\tau$ be a compactified $\Gamma_{\sigma}^\tau$
allowing $\sigma_{(1)}=\cdots=\sigma_{(\tau)}=0$.
For $\Psi\in\bar{\Gamma}_{\sigma}^\tau$, define the following continuous functions
$$
g_{\tau}(x;\Psi) =
  \sum_{k=1}^{\tau}\frac{\pi_{(k)}}{\sqrt{2}} \phi\bigg(\frac{x-\mu_{(k)}}{\sqrt{2}\epsilon_0}\bigg)
  +\sum_{k=\tau+1}^{p}\pi_{(k)}f(x;\theta_{(k)})
$$
where $f(x;\theta_{(k)})$ is density function of $k$th component.
Since $\sigma_{(p)}\geq\cdots\geq\sigma_{(\tau+1)}\geq\epsilon_0$,
$g_{\tau}(x;\Psi)$ is bounded over $\bar{\Gamma}_{\sigma}^\tau$.
Therefore, $\forall\Psi\in\bar{\Gamma}_{\sigma}^\tau$, we have
$\log E_{\Psi_0}\{g_{\tau}(X;\Psi)/f(X;\Psi_0)\}=-\Delta_{\tau}(\epsilon_0)<0$.
It is also obvious that $\Delta_{\tau}(\epsilon_0)$ is a decreasing function and $\underset{\epsilon_0\rightarrow0}{\lim}\Delta_{\tau}(\epsilon_0)\in(0,\infty)$.
Hence, the inequality $8\tau M\epsilon_0\log^2\epsilon_0<\Delta_{\tau}(\epsilon_0)$
holds for small enough $\epsilon_0$.

Define $l_n^{\tau}(\Psi)=\sum_{i=1}^n \log\{ g_{\tau}(x_i;\Psi) \}$
on $\bar{\Gamma}_{\sigma}^\tau$,
by the strong law of large numbers and
the upper bound of Jensen's inequality, we have almost surely
\begin{equation}\label{ld}
\underset{\Psi\in\bar{\Gamma}_{\sigma}^\tau}{\sup}
n^{-1}\{ l_n^{\tau}(\Psi)-\ell_n(\Psi_0) \}
\rightarrow E_{\Psi_0}\log\{ g_{\tau}(X;\Psi)/f(X;\Psi_0) \}
\leq -\Delta_{\tau}(\epsilon_0)
\end{equation}

For $\Psi\in\Gamma_{\sigma}^\tau$ and $\tau\in\{ 1,\cdots,p-1 \}$,
recall the definition of $A_{(k)},k\in\{1,\cdots,\tau\}$,
the mixture density $f(x_i;\Psi)\leq\frac{1}{\sigma_{(k)}}g_{\tau}(x_i;\Psi)$
for all $i\in A_{(k)}$.
While to the remaining observations,
since $|x_i-\mu_{(k)}|\geq|\sigma_{(k)}\log\sigma_{(k)}|$,
and if $\sigma_{(k)}$ is small enough that
$\sigma_{(k)}^{-1}=\exp\{ -\log\sigma_{(k)} \}< \exp\{ \frac{1}{4}\log^2\sigma_{(k)} \}$,
thus
$$
f(x;\theta_{(k)})\leq \frac{2}{\sigma_{(k)}}\phi\bigg(\frac{x-\mu_{(k)}}{\sigma_{(k)}}\bigg)
\leq \frac{1}{\sqrt{2}} \phi\bigg(\frac{x-\mu_{(k)}}{2\sigma_{(k)}}\bigg)
\leq \frac{1}{\sqrt{2}} \phi\bigg(\frac{x-\mu_{(k)}}{2\epsilon_0}\bigg)
$$
holds with $\sigma_{(k)}\leq\epsilon_0$, which implies $f(x_i;\Psi)\leq g_{\tau}(x_i;\Psi)$.

In summary, the log-likelihood contribution of $x_i$ have following upper bounds
$$
\log f(x_i;\Psi)\leq \bigg\{
\begin{array}{cc}
  -\log\sigma_{(k)}+\log g_{\tau}(x_i;\Psi), & i\in A_{(k)}, \\
  \log g_{\tau}(x_i;\Psi),                   & \textrm{otherwise}.
    \end{array}
$$
This further indicates the upper bound of log-likelihood
$$
\ell_n(\Psi)\leq l_n^{\tau}(\Psi) -\sum_{k=1}^\tau n(A_{(k)})\log\sigma_{(k)}.
$$

With the conclusions of (\ref{lp}) and (\ref{ld}), it can be show that
\begin{align*}
\underset{\Gamma_{\sigma}^\tau}{\sup} &p\ell_n(\Psi)-p\ell_n(\Psi_0) \\
&\leq
\underset{\Gamma_{\sigma}^\tau}{\sup} \{ l_n^{\tau}(\Psi)-\ell_n(\Psi_0) \}
+\underset{\Gamma_{\sigma}^\tau}{\sup}\sum_{k=1}^\tau \{
- n(A_{(k)})\log\sigma_{(k)}+p_n(\sigma_{(k)}) \} +o(n)\\
&\leq-n\Delta_{\tau}(\epsilon_0) + 4\tau Mn\epsilon_0\log^2\epsilon_0+o(n)
\leq -\frac{\Delta_{\tau}(\epsilon_0)}{2}n+o(n)
\end{align*}
for the chosen $\epsilon_0$. Note that $\Delta_{\tau}(\epsilon_0)>0$,
thus $\forall\tau\in\{ 1,\cdots,p-1 \}$,
$\sup_{\Gamma_{\sigma}^\tau}p\ell_n(\Psi)-p\ell_n(\Psi_0)\rightarrow-\infty$ a.s. as $n\rightarrow\infty$.
\end{proof}

\subsection*{Proof of Theorem \ref{thm3}}
\begin{proof}
When $\Psi\in\Gamma^c_{\sigma}\cap\Gamma_{\lambda}$,
since the component deviances have a positive lower bound and
divergent skew parameters do not lead to infinite component density,
$f(x;\Psi)$ is therefore bounded over $\Gamma^c_{\sigma}\cap\Gamma_{\lambda}$.

According to Jensen's inequality, we have
$E_{\Psi_0}\log\{ f(X;\Psi)/f(X;\Psi_0)\}<0$ for any
$\Psi\in\Gamma^c_{\sigma}\cap\Gamma_{\lambda}$.
We can also choose $\eta_0$ large enough so that
$\Psi_0\notin\Gamma^c_{\sigma}\cap\Gamma_{\lambda}$.
Consequently it is easy to show that, as in Wald (1949),
\begin{equation}\label{ll2}
\underset{\Gamma^c_{\sigma}\cap\Gamma_{\lambda}}{\sup}\bigg\{
\frac{1}{n}\sum_{i=1}^n\log\bigg( \frac{f(x_i;\Psi)}{f(x_i;\Psi_0)}\bigg)\bigg\}
\rightarrow -\Delta(\eta_0) <0 \ \ \textrm{a.s.} \ \ \textrm{as} \ n\rightarrow\infty.
\end{equation}

Note that $\Delta(\eta_0)$ is greater than zero and
is a increasing function of $\eta_0$.
With the upper bound in (\ref{ll2}) and the conditions
$\mathbf{C1}$-$\mathbf{C3}$, we get
\begin{align*}
\sup_{\Gamma^c_{\sigma}\cap\Gamma_{\lambda}} p\ell_n(\Psi) -p\ell_n(\Psi_0) &=
\underset{\Gamma^c_{\sigma}\cap\Gamma_{\lambda}}{\sup}
\sum_{i=1}^n\log\bigg( \frac{f(x_i;\Psi)}{f(x_i;\Psi_0)}\bigg) +
\sup_{\Gamma^c_{\sigma}\cap\Gamma_{\lambda}} p_n(\Psi)-p_n(\Psi_0)\\
&\leq -\frac{\Delta(\eta_0)}{2}n+o(n)
\end{align*}

Thus we have
$\sup_{\Gamma^c_{\sigma}\cap\Gamma_{\lambda}} p\ell_n(\Psi)-p\ell_n(\Psi_0)\rightarrow-\infty$
almost surely as $n\rightarrow\infty$.
\end{proof}

\subsection*{Proof of Theorem \ref{thm5}}
\begin{proof}
Based on the proof when $p=p_0$,
we establish a brief proof process for the case $p>p_0$.
With the defined distance (\ref{dst}) and any $\kappa>0$,
let us define a new parameter space
$\Omega(\kappa)=\{ \Psi: \Psi\in\Gamma, D(\Psi,\Psi_0)\geq \kappa \}$.
Clearly, $\Psi_0\notin\Omega(\kappa)$ when $\kappa>0$.

For $\Psi\in\Gamma_{\sigma}^p\cap\Omega(\kappa)$,
it is easy to show the derivations of Theorem \ref{thm1} are still applicable
by replacing $\Psi\in\Gamma_{\sigma}^p$ with
$\Psi\in\Gamma_{\sigma}^p\cap\Omega(\kappa)$.
Hence, we can quickly get
$\sup_{\Gamma_{\sigma}^p\cap\Omega(\kappa)} p\ell_n(\Psi)-p\ell_n(\Psi_0)\rightarrow -\infty$
as $n\rightarrow \infty$, and claim that
$\tilde{\Psi}\notin\Gamma_{\sigma}^p\cap\Omega(\kappa)$ with probability one.

Since $\Psi_0\notin\Omega(\kappa)$,
for $\Psi\in\Gamma_{\sigma}^\tau\cap\Omega(\kappa)$ where $1\leq\tau\leq(p-1)$
and $\Psi\in\Gamma^c_{\sigma}\cap\Gamma_{\lambda}\cap\Omega(\kappa)$,
the corresponding inequalities $E_{\Psi_0}\log\{ g_{\tau}(X;\Psi)/f(X;\Psi_0)\}<0$ and
$E_{\Psi_0}\log\{ f(X;\Psi)/f(X;\Psi_0)\}<0$
still holds respectively.
Thus (\ref{ld}) and (\ref{ll2}) can be extended to
\begin{align*}
\underset{\Gamma_{\sigma}^\tau\cap\Omega(\kappa)}{\sup} &
n^{-1}\{ l_n^{\tau}(\Psi)-\ell_n(\Psi_0) \}
\leq -\Delta_{\tau}(\epsilon_0)<0, \\
\underset{\Gamma^c_{\sigma}\cap\Gamma_{\lambda}\cap\Omega(\kappa)}{\sup} &\bigg\{
\frac{1}{n}\sum_{i=1}^n\log\bigg( \frac{f(X_i;\Psi)}{f(X_i;\Psi_0)}\bigg)\bigg\}
\rightarrow -\Delta(\eta_0)<0.
\end{align*}
for the properly selected $\epsilon_0, \eta_0$ and well-defined $g_{\tau}(x;\Psi)$.
Based on these two results, with $n\rightarrow \infty$, we similarly get
$\sup_{\Gamma_{\sigma}^\tau\cap\Omega(\kappa)} p\ell_n(\Psi)-p\ell_n(\Psi_0)\rightarrow -\infty$
for $\tau\in\{1,\cdots,(p-1)\}$ and $\sup_{\Gamma^c_{\sigma}\cap\Gamma_{\lambda}\cap\Omega(\kappa)} p\ell_n(\Psi)-p\ell_n(\Psi_0)\rightarrow -\infty$.

From the previous results,
it is clear that the penalized maximum likelihood estimator
$\tilde{\Psi}$ must fall in $\Gamma^*\cup\Omega^c(\kappa)$ with probability one.
Given the arbitrariness of $\kappa$,
$\tilde{\Psi}\in\Omega^c(\kappa)$ implies that $D(\tilde{\Psi},\Psi_0)\rightarrow0$.
At the same time, $\tilde{\Psi}\in\Gamma^*$ also implies
$D(\tilde{\Psi},\Psi_0)\rightarrow0$ by Kiefer and Wolfowitz (1956).
Thus, the strong consistency of the penalized MLE is proved
under the case $p>p_0$.

\end{proof}

\end{document}